\documentclass[prb,twocolumn,showpacs]{revtex4}     
\usepackage{epsfig,amsmath} 

\begin{document}

\title{Possible broken inversion and time-reversal symmetry state of electrons in bilayer graphene} 

\author{Xin-Zhong Yan$^{1}$ and C. S. Ting$^2$}
\affiliation{$^{1}$Institute of Physics, Chinese Academy of Sciences, P.O. Box 603, 
Beijing 100190, China\\
$^{2}$Texas Center for Superconductivity, University of Houston, Houston, Texas 77204, USA}
 
\date{\today}
 
\begin{abstract}
With the two-band continuum model, we study the broken inversion and time-reversal symmetry state of electrons with finite-range repulsive interactions in bilayer graphene. In the state, there are overlapped loop currents in each layer. With the analytical solution to the mean-field Hamiltonian, we obtain the electronic spectra. The ground state is gapped. In the presence of the magnetic field $B$, the energy gap grows with increasing $B$, in excellent agreement with the experimental observation. Such an energy-gap behavior originates from the disappearance of a Landau level of $n$ = 0 and 1 states. The present result resolves explicitly the puzzle of the gap dependence of $B$.   
\end{abstract}

\pacs{73.22.Pr,71.10.-w,71.27.+a} 

\maketitle

{\it Introduction.}
Bilayer graphene (BLG) has attracted much attention because of its potential application to new electronic devices. \cite{Ohta,Oostinga,McCann,Castro} Since the flat energy bands are sensitive to the electron-electron interactions, the ground state of the interacting electrons in BLG is highly different from the noninteracting picture. Theories have predicted various gapped broken symmetry states, such as a ferroelectric-layer asymmetric state \cite{Min,Nandkishore,Zhang,Jung,MacDonald} or a quantum valley Hall (QVH) state, \cite{Zhang2} a layer-polarized antiferromagnetic (AF) state, \cite{Gorbar,Throckmorton} a quantum anomalous Hall (QAH) state, \cite{Jung,Nandkishore1,Zhang1} a quantum spin Hall (QSH) state, \cite{Jung,Zhang1} a charge density wave state, \cite{Dahal} a superconducting state in coexistence with antiferromagnetism (SAF),\cite{Milovanovic} and the gapless states. \cite{Vafek,Lemonik} Though the experimental observations on the ground state are controversial (except one experiment has found a gapless state\cite{Mayorov}), several experiments \cite{Weitz,Freitag,Velasco,Bao} have provided the evidence for the existence of the gapped state at the charge neutrality point. In particular, a recent experiment performed on high quality suspended BLG \cite{Velasco} has found that the ground state is gapped and the gap grows with increasing magnetic field $B$ as 
\begin{eqnarray}
E_{\rm gap} = \Delta_0+\sqrt{a^2B^2 + \Delta_0^2} \label {gap}
\end{eqnarray}  
with $\Delta_0 \approx$ 1 meV and $a \approx$ 5.5 meVT$^{-1}$. This gap behavior cannot be explained by the existing microscopic theories because they show that the gap is weakly dependent of the magnetic field. The gap growing with the magnetic field is a puzzle.   

In this paper, based on Varma's loop-current idea developed in studying the mechanism of superconductivity in cuprates, \cite{Varma} we study the broken inversion and time-reversal symmetry state (BITRSS) of electrons with finite-range repulsive interactions in BLG. With the mean-field approximation (MFA) to the interacting electrons in a two-band continuum model (2BCM), we analytically solve the eigenstates of the electrons and prove Eq. (\ref{gap}) for the gap behavior in the presence of the magnetic field. 

{\it The 2BCM.}
We first briefly review the 2BCM. We start with the lattice structure of a BLG as shown in Fig. 1. The unit cell of BLG contains four atoms denoted as A, B on top layer, and A$^{\prime}$ and B$^{\prime}$ on bottom layer. The lattice constant defined as the distance between the nearest-neighbor (NN) A atoms is $a \approx 2.4$ \AA~. The energy of intralayer NN [between A (A$^{\prime}$) and B (B$^{\prime}$)] and interlayer NN (between B and A$^{\prime}$) electron hopping are $t \approx$ 2.9 eV and $t_1 \approx$ 0.31 eV, respectively. The Hamiltonian describing the noninteracting electrons is given by
\begin{eqnarray}
H=\sum_{k\sigma}C^{\dagger}_{k\sigma}H_kC_{k\sigma} \label {hm}
\end{eqnarray}
where $C^{\dagger}_{k\sigma}=(c^{\dagger}_{Ak\sigma},c^{\dagger}_{B'k\sigma},c^{\dagger}_{Bk\sigma },c^{\dagger}_{A'k\sigma})$ (in which $c^{\dagger}_{Ak\sigma}$ creates an electron at A sublattice with momentum $k$ and spin $\sigma$), and $H_k$ is given by  
\begin{eqnarray}
H_k = \begin{pmatrix}
0& F^{\dagger}\\
F&D\\
\end{pmatrix}\nonumber
\end{eqnarray}
with 0 as the 2$\times$2 zero matrix, and
\begin{eqnarray}
F = \begin{pmatrix}
e_k^{\ast}& 0\\
0&e_k\\
\end{pmatrix}, {\hskip 45mm}\nonumber\\
D = \begin{pmatrix}
0& -t_1\\
-t_1&0\\
\end{pmatrix},{\hskip 40mm}\nonumber\\
e_k = -t\left[2\cos\frac{k_x}{2}+\exp\left(-i\frac{\sqrt{3}k_y}{2}\right)\right]\exp\left(i\frac{k_y}{2\sqrt{3}}\right), \nonumber
\end{eqnarray}
and with the momentum $k=(k_x,k_y)$ (in unit of $a$ =1) confined to the first Brillouin zone. Note that $e_k$ vanishes at the Dirac points $K = (4\pi/3,0)$ and $K' = -K = (-4\pi/3,0)$. At $\pm K$, $e_k \approx \epsilon_0(\pm k_x+ik_y)$ where $\epsilon_0 = \sqrt{3}t/2$ and $k=(k_x,k_y)$ is measured from the Dirac point $K$ ($K'$) and confined to a small region of $K$ ($K'$). We hereafter use the unit of $\epsilon_0$ = 1 for energy. For the carrier concentration close to the charge neutrality point, we are concerned with the states close to the zero energy. Taking the transformation $C_{k\sigma}=T_kV_k$ with
\begin{eqnarray}
T_k = \begin{pmatrix}
1& 1\\
-D^{-1}F&1\\
\end{pmatrix},\nonumber
\end{eqnarray}
and 1 as the 2$\times$2 identity matrix, we have 
\begin{eqnarray}
T_k^{\dagger}H_kT_k = \begin{pmatrix}
h_k& 0\\
0&D\\
\end{pmatrix}, \nonumber\\
h_k = \begin{pmatrix}
0& e_k^2/t_1\\
e_k^{2\dagger}/t_1&0\\
\end{pmatrix}.
\end{eqnarray}
Clearly, $h_k$ describes the valence and conduction bands close to the zero energy, while the eigenvalue of $D$ is $\pm t_1$. The states connected by $D$ belong to the bands of overall energy separation $\pm t_1$ from the zero energy. For low-carrier doping, $h_k$ is the only part to be considered, which is the 2BCM \cite{McCann2,Novoselov} for noninteracting electrons. Within the 2BCM, the A and B$^{\prime}$ sublattices are considered for top and bottom layers, respectively.

\begin{figure} 
\centerline{\epsfig{file=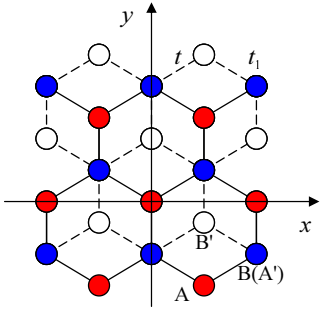,width=5. cm}}
\caption{(color online) Top view of BLG. Atoms A (A$^{\prime}$) and B (B$^{\prime}$) are on the top (bottom) layer. The parameters $t$ and $t_1$ are the electron hopping energies between the NN atoms belonging to the same layer, and to the neighboring layer above or below, respectively. } 
\end{figure} 

{\it The BITRSS.}
We here consider the interelectronic interaction effect. The interaction Hamiltonian is given as
\begin{eqnarray}
H' = \frac{1}{2}\sum_{ijll'}v_{li,l'j}\delta n_{li}\delta n_{l'j} \label {int}
\end{eqnarray}  
where $\delta n_{li}$ is the electron number deviation from the average occupation at site $i$ on layer $l$ (hereafter denoted as $li$ for short), and $v_{li,l'j}$ is the interaction between electrons at sites $li$ and $l'j$. As long as the exchange effect is considered, the exchange interactions between electrons are finite-ranged (or short-ranged) because the bare Coulomb interactions are screened due to the electronic charge density fluctuations. \cite{Yan} The operator product in the interaction Hamiltonian can be written as
\begin{eqnarray}
n_{li\sigma}n_{l'j\sigma}|_{li\ne l'j} = -2J^{\sigma}_{li,l'j}J^{\sigma}_{li,l'j}+(n_{li\sigma}+n_{l'j\sigma})/2 \nonumber
\end{eqnarray}  
where $n_{li\sigma}$ is the density operator of spin $\sigma$, and $J^{\sigma}_{li,l'j}= (c^{\dagger}_{li\sigma}c_{l'j\sigma}-c^{\dagger}_{l'j\sigma}c_{li\sigma})/2i$ is proportional to the current operator. Therefore, the electrons can be coupled through the currents. Using the MFA and considering only the current couplings, we get
\begin{eqnarray}
H' = 2i\sum_{ll'ij\sigma}v_{li,l'j}\langle J^{\sigma}_{li,l'j}\rangle c^{\dagger}_{li\sigma} c_{l'j\sigma}. \label {mfa1}
\end{eqnarray} 
Here, $\langle J^{\sigma}_{li,l'j}\rangle$ is a statistical average to be self-consistently determined with the mean-field theory. For simply exploring the physics, in this work, we consider the case that the average is nonvanishing only for the intralayer bonds and denote the average as $\langle J^{\sigma}_{li,l'j}\rangle = \pm J_l(|\vec j-\vec i|)\delta_{ll'}$. Here, $J_l(|\vec j-\vec i|)$ means a bond current between the two sites $li$ and $lj$. The sign factor $\pm$ depends on the direction of the electron motion from site $i$ to site $j$. Under the assumption of broken inversion symmetry, we have $J_1$ (of top layer) = -$J_2$ (of bottom layer).

\begin{figure} 
\centerline{\epsfig{file=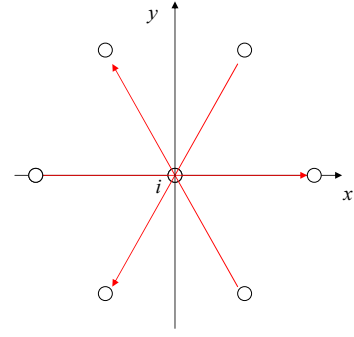,width=5. cm}}
\caption{(color online) Current $\langle J^{\sigma}_{li,lj}\rangle = \pm J_l$ between site $i$ and its NNs $j$ on a layer. The + (-) sign is for the electron motion along (opposite to) an arrow.} 
\end{figure} 

For the sake of illustration, we first consider the simplest case that there exist only the NN bond currents. We denote the NN interaction as $v_{li,lj} = v$ and the NN bond current as $J_l(|\vec j-\vec i|) = J_l$. Figure 2 shows the picture of the bond currents between the site $i$ and its NN sites. The sign + (-) is for the motion along (opposite to) an arrow. The total current density passing through the site $i$ is zero. For every site on a layer, one can depict such a figure. On the triangle lattice of the top or bottom layer, one then can find the triangle current loops surrounding every three NN sites. Therefore, Fig. 2 shows an equivalent view of the loop currents on a layer. To proceed, we rewrite Eq. (\ref{mfa1}) for the present case in momentum space,
\begin{eqnarray}
H' = \sum_{kl\sigma}\Delta_l(k) c^{\dagger}_{lk\sigma} c_{lk\sigma}, \label {mfa2}
\end{eqnarray}
with $\Delta_l(k) = J_lv[8\sin(k_x/2)\cos(\sqrt{3}k_y/2)-4\sin k_x] \equiv J_lv\xi(k)$ [$\xi(k)$ is so defined for later use]. Since the low-energy states are considered as aforementioned, we take the leading term of the expansion of $\Delta_l(k)$ at the Dirac points $\pm K$,
\begin{eqnarray}
\Delta_l(\pm K) = \pm 6\sqrt{3}J_lv \equiv s_vs_l\Delta  \label{op}
\end{eqnarray}
where $s_v = \pm 1$ for valley $\pm K$, $s_l$ = 1 (-1) for top (bottom) layer, and $\Delta = 6\sqrt{3}J_1v$. The quantity $\Delta$ is the order parameter of the BITRSS. The signs $s_v$ and $s_l$ reflect the broking of time-reversal and inversion symmetries, respectively. Equation (\ref{op}) is an important result. 

Now the effective 2BCM Hamiltonian is obtained as 
\begin{eqnarray}
\tilde h_{vk} = \begin{pmatrix}
s_v\Delta & \epsilon_{vk}\\
\epsilon^{\dagger}_{vk}&-s_v\Delta\\
\end{pmatrix} \label {heff}
\end{eqnarray}
with $\epsilon_{vk} = (s_vk_x+ik_y)^2/t_1$. The eigenvalues are $\pm \sqrt{\Delta^2 +|\epsilon_{vk}|^2} \equiv \pm E_k$ and the corresponding eigen-wave-functions $\psi_{vk}$ are given by
\begin{eqnarray}
\psi_{vk} = \begin{pmatrix}
\pm R^{\pm}_{vk}\\
R^{\mp}_{vk}e^{-i\phi_{vk}}\\
\end{pmatrix}, \nonumber
\end{eqnarray}
where $R^{\pm}_{vk} = \sqrt{1\pm s_v\Delta/E_k}/\sqrt{2}$, and $\phi_{vk} = \arg(\epsilon_{vk})$. For the system at the charge neutrality point, the order parameter $\Delta$ is determined by
\begin{eqnarray}
\Delta &=& -\frac{\sqrt{3}v}{2N}\sum_{k}\xi(k)\langle c^{\dagger}_{Ak\sigma}c_{Ak\sigma}\rangle \nonumber\\
&=& \frac{9v}{N}{\sum_k}'\left(\langle c^{\dagger}_{Ak\sigma}c_{Ak\sigma}\rangle_{K'}-\langle c^{\dagger}_{Ak\sigma}c_{Ak\sigma}\rangle_K\right) \nonumber\\
&=& \frac{9\sqrt{3}v\Delta}{2V}{\sum_k}'(f^-_k-f^+_k)/E_k
\label {od}
\end{eqnarray}
where the first line is the definition with the $k$ summation runs over the first Brillouin zone [with $N$ the total number of A (B$^{\prime}$) atoms on top (bottom) layer], the summation in the second line is separated into two parts over valleys $K$ and $K'$ with $\xi(k) \approx \xi(\pm K) = \pm 6\sqrt{3}$ being used, and in the last line the averages are carried out using the eigen-wave-functions given above. Here $f^{\pm}_k = f(\pm E_k)$ with $f(E)$ as the Fermi distribution function, and $V = \sqrt{3}N/2$ is the single-layer area (in unit of $a$ = 1). For the ground state, we can integrate out Eq. (\ref{od}) and get
\begin{eqnarray}
\frac{9\sqrt{3}t_1v}{8\pi}\ln\left(\frac{k_c^2}{t_1|\Delta|}+\sqrt{1+\frac{k_c^4}{t^2_1\Delta^2}}\right) = 1 \label {od0}
\end{eqnarray}
where $k_c \approx t_1$ is the momentum cutoff. 

\begin{figure} 
\centerline{\epsfig{file=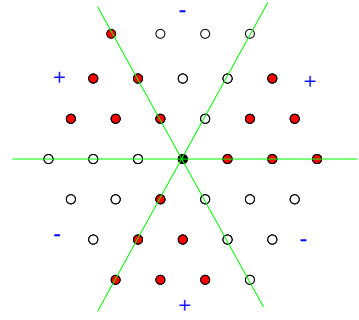,width=5. cm}}
\caption{(color online) Phase for the current $\langle J^{\sigma}_{li,lj}\rangle = \pm J_l(|\vec i-\vec j|)$. The sign + (-) is for the electron motion from site $i$ as the center point to a red (white) site $j$.} 
\end{figure} 

To reproduce the experimental data $|\Delta|$ = 1 meV from Eq. (\ref{od}), $v \approx 2\epsilon_0$ is needed. Unfortunately, this value of $v$ is too strong to acquire for electrons in graphene. Using $a$ = 2.4 \AA~ and the dielectric constant of graphene $\sim 4$, the typical value of $v$ is obtained as  $0.6\epsilon_0$. The problem stems from considering only the NN bond currents. If we include the contribution from the currents between long bond sites, the strength of order parameter $\Delta$ = 1 meV can be obtained. To repeat the MFA for the case of finite-range interactions, we note that $\langle J^{\sigma}_{li,lj}\rangle = \pm J_l(|\vec j-\vec i|)$ depends on the vector $\vec d = \vec j-\vec i$. The sign factor $\pm$ of all the currents passing through the site $i$ is illustrated in Fig. 3. In this case, there are various overlapped triangle current loops in each layer. By the MFA, we obtain the similar equation for determining the order parameter as Eq. (\ref{op}) but with $v$ replaced with
\begin{eqnarray}
v_{\rm eff} = \frac{2}{9}\sum_{\vec d}v(d)\sin^2(K_xd_x)  \label{veff}
\end{eqnarray}
where $v(d) = v_{li,lj}$, and $K_x = 4\pi/3$. Here the $\vec d$ summation runs over the sites of A (or B$^{\prime}$) sublattice. If the summation runs over only the six NNs, $v_{\rm eff}$ reduces to $v$. For the general case of finite-range repulsive interactions, one has $v_{\rm eff} > v$.

{\it Landau states.}
In the presence of a magnetic field $B$ applied perpendicularly to the sample plane, we take the Landau gauge for the vector potential, $A = (0, Bx)$. In this gauge, the $y$ component momentum $k_y$ is a good quantum number. Replacing the variable $x$ and the operator $k_x = -i\nabla_x$ with the raising and lowering operators $a^{\dagger}$ and $a$, $k_y+Bx = \sqrt{B/2}(a^{\dagger}+a)$ and $k_x = i\sqrt{B/2}(a^{\dagger}-a)$, we get from Eq. (\ref{heff})
\begin{eqnarray}
\tilde h_{K} = \begin{pmatrix}
\Delta & -\omega_ca^{\dagger 2}\\
-\omega_ca^2&-\Delta\\
\end{pmatrix} \label {hmK}
\end{eqnarray}
for the quasiparticles at valley $K$ and
\begin{eqnarray}
\tilde h_{K'} = \begin{pmatrix}
-\Delta & -\omega_ca^2\\
-\omega_ca^{\dagger 2}&\Delta\\
\end{pmatrix} \label {hmK'}
\end{eqnarray}
for the quasiparticles at valley $K'$, and $\omega_c = 2B/t_1$ is the cyclotron frequency. The eigenvalue $e_n$ and eigenvector $\psi_n$ of $\tilde h_K$, for $n \geq 2$, are given by
\begin{eqnarray}
e_n = E_n, ~~~~
\psi_n = \begin{pmatrix}
R^+_n\phi_{n}\\
-R^-_n\phi_{n-2}\\
\end{pmatrix}\nonumber\\
e_n = -E_n, ~~~~
\psi_n = \begin{pmatrix}
R^-_n\phi_{n}\\
R^+n\phi_{n-2}\\
\end{pmatrix}\nonumber 
\end{eqnarray}
where $E_n = \sqrt{\Delta^2 + \omega_c^2n(n-1)}$, $R^{\pm}_n = \sqrt{1\pm \Delta/E_n}/\sqrt{2}$, and $\phi_n$ is the $n$th level wave function of a harmonic oscillator of mass $m = t_1/2$ and frequency $\omega_c$ centered at $x_c = -k_y/B$. For $n$ = 0 and 1, we have \begin{eqnarray}
e_{0,1} = \Delta, ~~~~
\psi_{0,1} = \begin{pmatrix}
\phi_{0,1}\\
0\\
\end{pmatrix}.\nonumber
\end{eqnarray}
While at valley $K'$, the solutions are
\begin{eqnarray}
e_n &=& E_n, ~~~~
\psi_n = \begin{pmatrix}
R^-_n\phi_{n-2}\\
-R^+_n\phi_{n}\\
\end{pmatrix},~~~~n \geq 2 \nonumber\\
e_n &=& -E_n, ~~~~
\psi_n = \begin{pmatrix}
R^+_n\phi_{n-2}\\
R^-_n\phi_{n}\\
\end{pmatrix},~~~~n \geq 2\nonumber \\
e_{0,1} &=& \Delta, ~~~~
\psi_{0,1} = \begin{pmatrix}
0\\
\phi_{0,1}\\
\end{pmatrix}.\nonumber
\end{eqnarray}

We note that there is no state of energy $-\Delta$ for $n$ = 0 and 1. The particle-hole symmetry is no longer valid in the presence of the magnetic field, in qualitative agreement with the experimental observation. \cite{Velasco} The energy gap is $E_{\rm gap}= e_0-(-E_2)$ if $\Delta > 0$ or $E_2 - e_0$ if $\Delta < 0$. The gap can be expressed as 
\begin{eqnarray}
E_{\rm gap} = |\Delta|+\sqrt{2\omega_c^2 + \Delta^2} \label {gp}
\end{eqnarray}  
which is exactly the same form as Eq. (\ref{gap}). The factor $a$ in Eq. (\ref{gap}) corresponds to $a = 2\sqrt{2}/t_1$ (in units of $\epsilon_0 = 1$ and lattice constant $a$ = 1), which is independent of the strength of the interactions. From the absolute values of the parameters $t$, $t_1$ and $a$ given above, we get $a \approx 5.2$ meVT$^{-1}$, which is close to the experimental data $5.5$ meVT$^{-1}$.

Recently, Zhu {\it et al.} have studied the same problem by numerically diagonalizing a mean-field Hamiltonian on a lattice of finite size because their model cannot be analytically solved. \cite{Zhu} They have also reached the same conclusion that the gap grows with increasing $B$. 

Note that the present effective two-band Hamiltonian does not apply to the edge of the system because where the coordination number of each site is different from that in the bulk and some current loops disappear. Therefore, the form of the hopping term in momentum space and the self-energy are all different from that in the bulk. 

The ground state observed by the experiment \cite{Velasco} is insulating at the charge neutrality point without external electric and magnetic fields. Also, the gap can be closed by an electric field of either sign perpendicular to the graphene plane. The experimental observation thus rules out the SAF \cite{Milovanovic} as well as the gapless states. \cite{Vafek,Lemonik} The QAH and QSH states are excluded because the conductance $\sim 4e^2/h$ yielded by their edge states is not observed. The ferroelectric-layer asymmetric state \cite{Min,Nandkishore,Zhang,Jung,MacDonald} or the QVH state \cite{Zhang2} not only cannot produce the gap behavior with $B$, but also contradicts to the observation of no net charge polarization between the layers. The order parameter of the AF state \cite{Gorbar,Throckmorton} is weakly dependent of the magnetic field.\cite{note} Therefore, among the existing proposed states, the overlapped loop-current state is the most possible candidate for the ground state of electrons in BLG.

{\it Summary.}
We have studied the BITRSS of the electrons with finite-range repulsive interactions in BLG. With the 2BCM, we have analytically solved the eigenstates of the mean-field Hamiltonian and obtained the gapped ground state at the charge neutrality point. The order parameter is odd for both interchanges of valleys and layers. In the presence of the magnetic field applied perpendicularly to the layers, we have obtained the spectra of the quasiparticles. Because of the breaking of time-reversal symmetry, a Landau level of $n$ = 0 and 1 states disappears, resulting in the large energy gap growing with increasing the magnetic field. The present result is in excellent agreement with the experimental observation. \cite{Velasco}

This work was supported by the National Basic Research 973 Program of China under Grants No. 2011CB932700 and No. 2012CB932300, NSFC under Grant No. 10834011, and the Robert A. Welch Foundation under Grant No. E-1146.


\begin{thebibliography}{99}

\bibitem{Ohta} T. Ohta, A. Bostwick, T. Seyller, K. Horn, E. Rotenberg, Science {\bf 313}, 951 (2006).

\bibitem{Oostinga} J. B. Oostinga, H. B. Heersche, X. Liu, A. F. Morpurgo, and L. M. K. Vanderspen, Nature Mater. {\bf 7}, 151 (2008).

\bibitem{McCann} E. McCann, Phys. Rev. B {\bf 74}, 161403(R) (2006).

\bibitem{Castro} E. V. Castro, K. S. Novoselov, S. V. Morozov, N. M. R. Peres, J. M. B. Lopes dos Santos, J. Nilsson, F. Guinea, A. K. Geim, and A. H. Castro Neto, Phys. Rev. Lett.  {\bf 99}, 216802 (2007).

\bibitem{Min} H. K. Min, G. Borghi, M. Polini, and A. H. MacDonald, Phys. Rev. B {\bf 77}, 041407(R) (2008).

\bibitem{Nandkishore} R. Nandkishore and L. Levitov, Phys. Rev. Lett. {\bf 104}, 156803 (2010).

\bibitem{Zhang} F. Zhang, H. K. Min, M. Polini, and A. H. MacDonald, Phys. Rev. B {\bf 81}, 041402(R) (2010).

\bibitem{Jung} J. Jung, F. Zhang, and A. H. MacDonald, Phys. Rev. B {\bf 83}, 115408 (2011).

\bibitem{MacDonald} A. H. MacDonald, J. Jung, and F. Zhang, Phys. Scr. {\bf T146}, 014012 (2012).

\bibitem{Zhang2} F. Zhang and A. H. MacDonald, Phys. Rev. Lett. {\bf 108}, 186804 (2012).

\bibitem{Gorbar} E. V. Gorbar, V. P. Gusynin, V. A. Miransky, and I. A. Shovkovy, Phys. Rev. B {\bf 85}, 235460 (2012).

\bibitem{Throckmorton} R. E. Throckmorton and O. Vafek, Phys. Rev. B {\bf 86}, 115447 (2012).

\bibitem{Nandkishore1} R. Nandkishore and L. Levitov, Phys. Rev. B {\bf 82}, 115124 (2010).

\bibitem{Zhang1} F. Zhang, J. Jung, G. A. Fiete, Q. Niu, and A. H. MacDonald, Phys. Rev. Lett. {\bf 106}, 156801 (2011).

\bibitem{Dahal} H. Dahal, T. Wehling, K. Bedell, J. Zhu, and A. V. Balatsky, Physica. {\bf 405}, 2241 (2010).

\bibitem{Milovanovic} M.V. Milovanovi\'c and S. Predin, Phys. Rev. B {\bf 86}, 195113 (2012).

\bibitem{Vafek} O. Vafek, and K. Yang, Phys. Rev. B {\bf 81}, 041401 (2010). 

\bibitem{Lemonik} Y. Lemonik, I. L. Aleiner, C. Toke, and V. I. Fal'ko, Phys. Rev. B {\bf 82}, 201408(R) (2010).

\bibitem{Mayorov} A. S. Mayorov, D. C. Elias, M. Mucha-Kruczynski, R. V. Gorbachev, T. Tudorovskiy, A. Zhukov, S. V. Morozov, M. I. Katsnelson, V. I. Fal\'ko, A. K. Geim, and K. S. Novoselov, Science {\bf 333}, 860 (2011).

\bibitem{Weitz} R. T. Weitz, M. T. Allen, B. E. Feldman, J. Martin, and A. Yacoby, Science {\bf 330}, 812 (2010).

\bibitem{Freitag} F. Freitag, J. Trbociv, M. Weiss, and C. Sch\"nenberger, Phys. Rev. Lett. {\bf 108}, 076602 (2012).

\bibitem{Velasco} J. Velasco Jr., L. Jing, W. Bao, Y. Lee, P. Kratz, V. Aji, M. Bockrath, C. N. Lau, C. Varma, R. Stillwell, D. Smirnov, F. Zhang, J. Jung, and A. H. MacDonald, Nature Nanotech. {\bf 7}, 156 (2012).

\bibitem{Bao} W. Bao, J. Velasco Jr., L. Jing, F. Zhang, B. Standley, D. Smirnov, M. Bockrath A. H. MacDonald, and C. N. Lau, Proc. Natl. Acad. Sci. USA {\bf 109}, 10802 (2012).

\bibitem{Varma} C. M. Varma, Phys. Rev. Lett. {\bf 83}, 3538 (1999).

\bibitem{McCann2} E. McCann and V. I. Fal'ko, Phys. Rev. Lett. {\bf 96}, 086805 (2006).

\bibitem{Novoselov} K. S. Novoselov, E. McCann, S. V. Morozov, V. I. Fal'ko, M. I. Katsnelson. U. Zeitler, D. Jiang, F. Schedin, and A. K. Geim, Nat. Phys.  {\bf 2}, 177 (2006).

\bibitem{Yan} X.-Z. Yan and C. S. Ting, Phys. Rev. B {\bf 86}, 125438 (2012).

\bibitem{Zhu} L. J. Zhu, V. Aji, and C. M. Varma, arXiv: 1202.0821.

\bibitem{note} The magnitude of the AF order parameter depends sensitively on the momentum cutoff in a 2BCM. With a more reasonable four-band continuum model than the 2BCM, it can be shown that the AF gap depends weakly on the magnetic filed. We will show the result in a future publication.

\end{thebibliography}
\end{document}